# Investigation of cloud cavitation passive control method using Cavitating-bubble Generators (CGs)

Ebrahim Kadivar*; Ould el Moctar

*University of Duisburg-Essen, Duisburg, Germany*

**Abstract**

In this article, we propose a passive method to control unsteady cloud cavitation on hydrofoils using cavitation-bubble generator (CGs). This method may be used in many engineering applications, in particular in marine and turbomachinery. First, we used a Partially-averaged Navier Stokes (PANS) model for turbulence to simulate the unsteady cavitating flow and validated it based on experimental data. This model was coupled with a mass transfer model and implemented to the open source software package OpenFOAM. Second, the effect of a proper design of CGs on qualitative parameters such as cavitation structure and the shape of cavity were studied. The effect of CGs on the destructive effects of cavitation such as vibration, turbulent velocity fluctuations and high-pressure amplitude were analyzed. Our results showed that a proper design of CGs may reduce the amplitude of the force fluctuations on the hydrofoil substantially. Further on, the local boundary layer around the hydrofoil surface was altered and the turbulent velocity fluctuation was reduced significantly using this technique.

**Introduction**

Unsteady cloud cavitation is a flow phenomenon that may have destructive effects such as pressure pulsation, noise and erosion [1] - [2]. The control of the unsteady cavitation is expected to improve the performance of blades in turbines, propellers and rudders specially in marine and turbomachinery applications. In the past decades, most of the researchers were focused on resolving cavitation phenomenon numerically and experimentally but how the destructive effects of cavitation can be controlled has been studied scare. Few investigations were conducted to control the cavitation and reduce its undesirable behavior in the supercavitation and cloud cavitation regimes. Crimi et al. [3] investigated the effect of introducing a sweep angle to a hydrofoil. They carried out tests on semi-span hydrofoils with sweep angles of 0, 15, 30 and 45 degrees. They showed that the sweep angle may alleviate the problem of erosion due to cavitation. Kuiper [4] improved the inception behavior of tip vortex cavitation on propellers. He added forward skew to the propeller blade and compared the result to those obtained from cases with no skew. Ausoni et al. [5] investigated the hydrofoil roughness effects on von karman vortex shedding. They showed that with the help of a distributed roughness, the transition to turbulence is triggered at the leading edge, which reduces the span-wise non uniformities in the boundary layer transition process. Xiang et al. [6] investigated the effect of the ventilated partial cavity on the drag reduction. Their results showed that a remarkable drag reduction may be achieved for the lower and higher cavitation cases. Kadivar et al. [7] investigated supercavitation flow over different 30, 45 and 60 degree wedge cavitators. They showed that the drag coefficient decreases by the reduction of the wedge angle of cavitator. They illustrated that the wedge angle of the cavitator affects significantly the shape and type of supercavitation. Coutier-Delgosha et al. [8] studied the effect of the surface roughness on the dynamics of sheet cavitation on a two-dimensional foil section. They presented that the roughness in the downstream end of the sheet cavity plays a major role in the arrangement of the cavitation cycle. According to the previous studies it is well known that the control of cavitation on the immersed bodies has a significant influence on the reduction of the destructive effects of cavitation such as loss of efficiency and vortex-induced vibration. We introduce in this work a method to control the unsteady cloud cavitation. The idea was adapted from vortex generators that are common in boundary layer control around airfoils in aerospace engineering applications. This analogy is used to control boundary layer and as a consequence the upper surface pressure distribution and to reduce the destructive effects of cavitation.

*Corresponding Author, Ebrahim Kadivar: ebrahim.kadivar@uni-due.de

**Numerical Modelling**

In this paper PANS model was used for the simulation of unsteady cavitating flow. The PANS model is a hybrid method of Reynolds-averaged Navier Stokes (RANS) and Direct Numerical Simulation (DNS), which was first proposed by Girimaji [9]. The results of other publications using this model showed the improvement of the accuracy of numerical simulation, e.g. Song and Park, [10] and Ji et al. [11]. This approach is coupled with a mass transfer model which was implemented to the open source software package OpenFOAM. The interface between the liquid and vapor phases is captured by using the volume of fluid (VOF) method and the Schnerr and Sauer [12] cavitation model is chosen for numerical simulation of cavitation.

**Passive cavitation control**

Cavitation bubbles are generated artificially to influence the whole processes of vaporization, bubble generation and bubble implosion, which occur at normal condition without any control. This is done by inserting a type of micro vortex generators called cavitating-bubble generators (CGs) on the upper surface of the hydrofoil where it is expected that the cavitation is produced naturally. The 2D schematic view of a hydrofoil with CGs on the suction side of the hydrofoil and the 3D view of CGs located on the hydrofoil are shown in Figure 1. "L" and "H" are the dimensionless length and height of the CGs with regard to the chord length of the hydrofoil. Our investigations on the size of the CGs show that it should be small enough so that it does not have a significant effect on the hydrodynamics performance of the hydrofoil. Because of unpleasant side effects that may occur, the shape, the size and the location of the CGs are curtail.

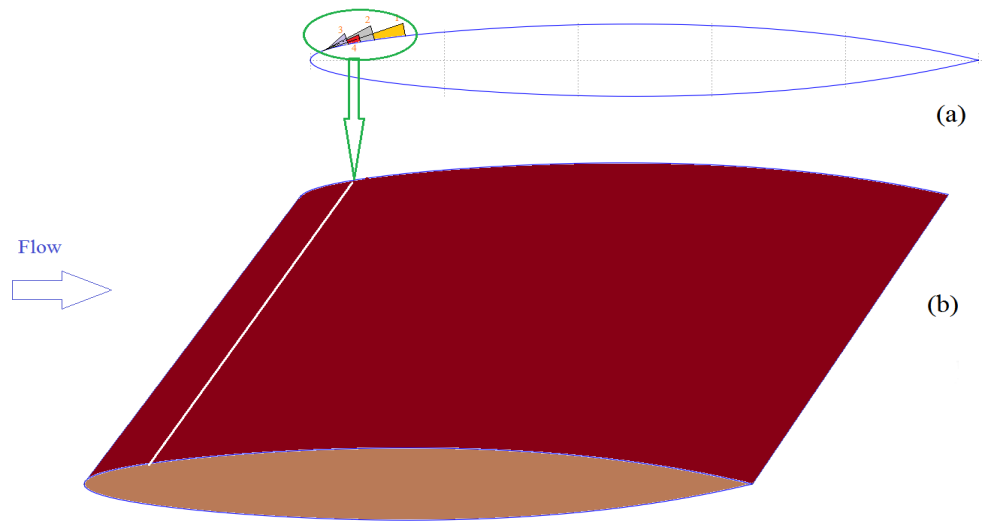

Figure 1. a) 2D view of some example CGs located on the hydrofoil with different sizes and angles, b) 3D view of a spanwise CG located on the upstream of the hydrofoil suction side

**Results and discussion**

Results of unsteady cavitating flow over the CAV2003 benchmark hydrofoil with and without of the cavitation control are presented here. The geometry of this hydrofoil was perfected at the CAV2003 workshop, [13]. To confirm the results independency on the mesh size and on the time step value these two parameters are tested in this work. The effect of using three different mesh sizes on time-averaged lift and drag coefficients and Strouhal number based on the chord $St_c = f \times l_{ref}/V_{ref}$ are shown in Table. 1, which f and $V_{ref}$ are the cavity self-oscillation frequency and reference velocity respectively. The Grids A to C have 1,150,000, 3,450,000 and 7,257,600 cells respectively. At first the time step for the unsteady computation was set to 4e-6. The results show that the changes between the last two grids are small. Therefore, we performed our simulations using grid B in the entire computational domain.

*Corresponding Author, Ebrahim Kadivar: ebrahim.kadivar@uni-due.de

|  | Time-averaged $C_l$ | Time-averaged $C_d$ | $St_c$ |
|---|---|---|---|
| Grid A | 0.48 | 0.075 | 0.107 |
| Grid B | 0.44 | 0.073 | 0.11 |
| Grid C | 0.43 | 0.072 | 0.11 |
| Delgosha-Simulation | 0.45 | 0.07 | 0.108 |
| Delgosha-Experiment | - | - | 0.15 |

Table 1. Comparison of the time-averaged lift and drag coefficients and Strouhal number based on the chord at cavitation number σ = 0.8

After that for grid B we have simulated with two different time steps 4e-6 s and 1e-5 s. The results showed that the difference between these two time steps is not significant. Time history of lift $C_l$ and drag coefficients $C_d$ from our numerical simulation were compared with the work by Delgosha et al. as shown in Fig. 2.

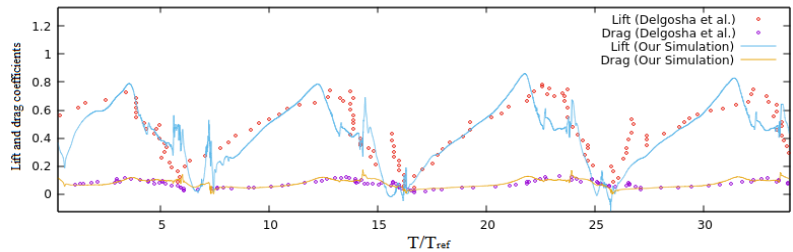

Figure 2. Time history of lift and drag coefficients, present work and Delgosha et al. (2007)

Figure 3 shows that the amplitude of the force fluctuations using proper CGs was reduced significantly and the cavitating flow reached a quasi stable situation. This figure shows that using proper size and position of CGs on the hydrofoil surface the advantage of reduction of the pressure drag force around the hydrofoil will be obtained. The pressure drag was reduced more in comparison with the lift reduction. That means with the proper design of passive controller the hydrodynamic efficiency was increased and the unsteady behavior of the cavitation was suppressed. Figure 4 and 5 present the pressure distributions around the hydrofoil with and without using passive controller in one typical cycle.

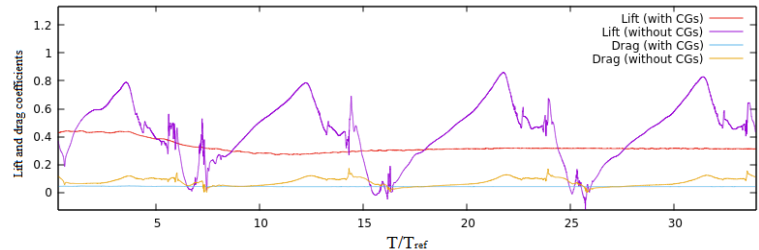

Figure 3. Time history of lift and drag coefficients with a proper CG on the hydrofoil and the hydrofoil without using CGs.

When the large number of vapor structures as bubble clusters and small-scale vortices reach the high pressure region on the surface of hydrofoil, they collapse. This process may induce different destructive effects such as high pressure peaks on the solid surface of hydrofoil and wall-pressure fluctuations. As be seen in Figure 4 for the hydrofoil without CGs high pressure peaks are captured during the cavitation process. It can be seen from the instantaneous figures that these pressure peaks were occurred near the rear of attached cavity at the cavity collapse region and at the trailing edge of the hydrofoil. Figure 5 shows that for the hydrofoil with proper CGs no pressure peaks are observed. The high peak values of the pressure using proper design of CGs were reduced significantly and the cavitating flow reaches a quasi-steady state situation which shows no periodical large cloud cavitation.

*Corresponding Author, Ebrahim Kadivar: ebrahim.kadivar@uni-due.de

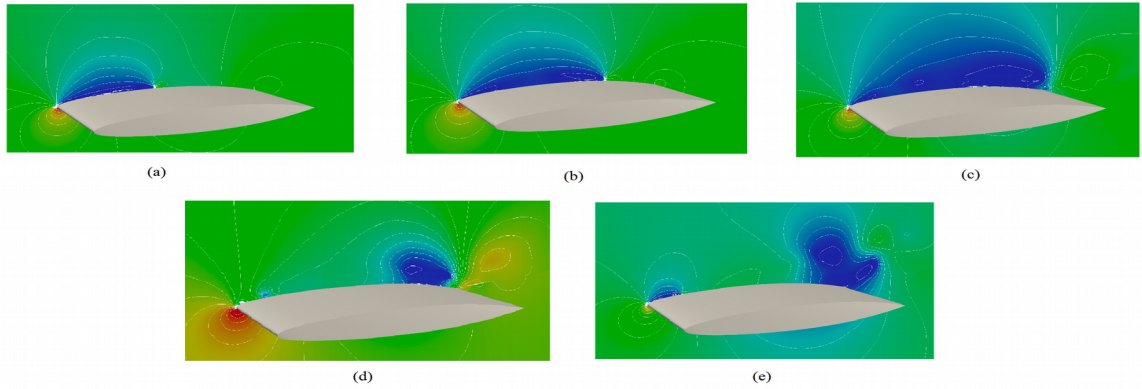
Figure 4. Pressure distributions at the middle of spanwise of the hydrofoil without CGs in one typical cycle.

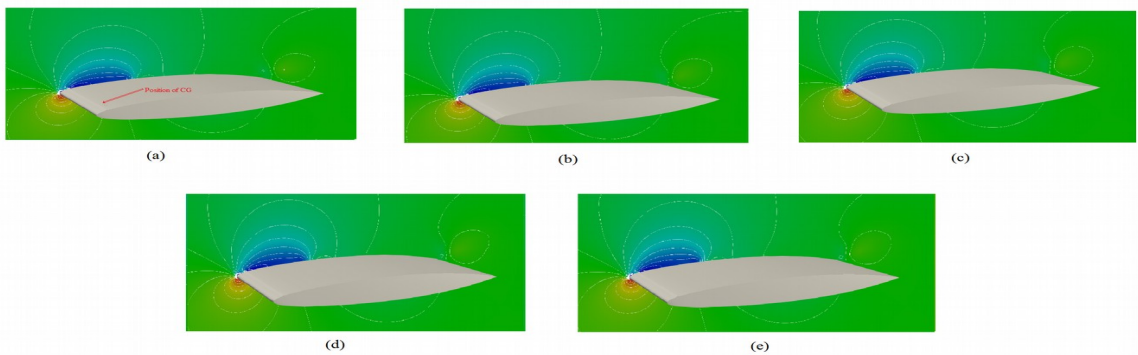
Figure 5. Pressure distributions at the middle of spanwise of the hydrofoil with using a proper CGs in one typical cycle.

For the case without CGs four different peak frequencies $f_1$ = 6.6 Hz, $f_2$ = 12.4 Hz, $f_3$ = 19.13 Hz and $f_4$ = 25.8 Hz and for the case with CGs two different peak frequencies $f_1$ = 10.23 Hz and $f_2$ = 16.8 Hz were observed. These values show that the dominant frequency for the case of using CGs is increased insignificantly in comparison with the simple one. However, the amplitude of the dominant frequency using proper CGs was reduced remarkably. The dominant frequency with the highest amplitude corresponding to the cavitation shedding events is considered to find the Strouhal number. For the case of ith CGs the dominant frequency corresponds to Strouhal number of $St_c$ = 0.17 and $St_l$ = 0.093, while for the case without CGs the Strouhal number values are about $St_c$ = 0.11 and $St_l$ = 0.077. Since the unsteady cloud cavitation and the shedding of sheet cavity this are complex phenomena, Q-criterion reflected the structure of the cloud cavitation and vorticity distribution properly. This criterion defined as the positive second invariant of the velocity gradient tensor, [14]. Figure 6 and 7 show the dynamics of the vortex structure evolution as visualized by the Q-criterion during one oscillation cycle with and without using a proper manner of passive controller. The vortex structures and the sudden changes of cavity cause to the significant changes at the downstream of the sheet cavitation, which induce a strong variation of lift and drag forces on the hydrofoil surface. The figures show that the sheet cavity shedding and the vapor cloud cavity have a substantial influence on the vortex structures. The main mechanism for the generation of the cavitating horse-shoe vortex could be the interaction between the circulating flow and the shedding of vapor cloud. It can be seen from Figure 7 (a-h) that the local flow structures and the local boundary layer don't change significantly. They show a quasi-constant turbulent velocity process in the cavitation regimes around the hydrofoil. According to instantaneous images, it is clear that the local boundary layer and wake flow changes insignificantly under the effect of CGs on the hydrofoil surface which means that the vortex structure of the cavity in the closure and wake regions changed remarkably in comparison with the hydrofoil without CGs.

*Corresponding Author, Ebrahim Kadivar: ebrahim.kadivar@uni-due.de

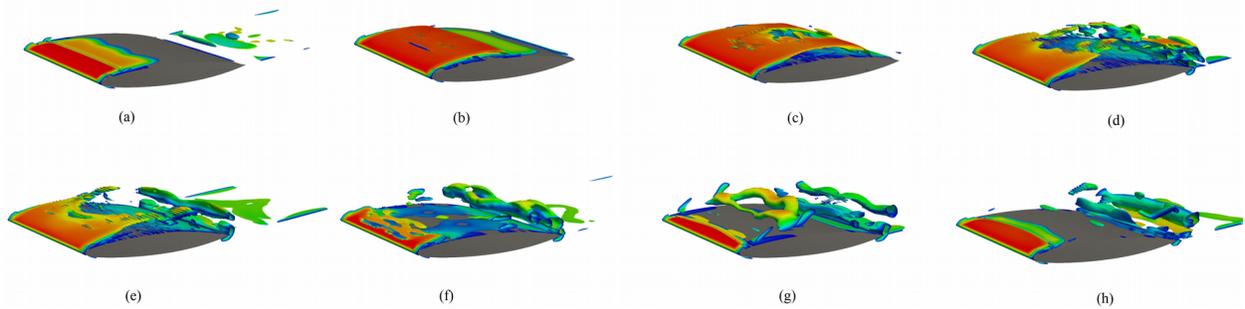
Figure 6. Instantanous Iso-surface of the Q-criterion with 100,000 [s$^{-2}$]) colored by streamwise velocity without using CGs.

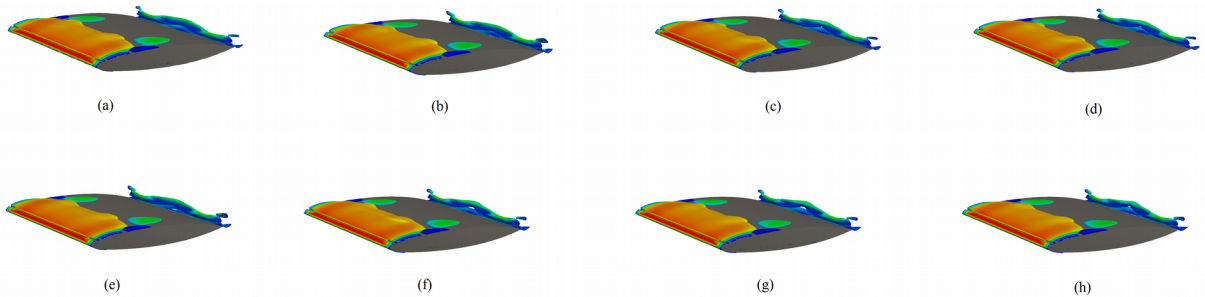
Figure 7. Instantanous Iso-surface of the Q-criterion with 100,000 [s$^{-2}$]) colored by streamwise velocity with using CGs.

## Conclusions

In the present study, a method to control unsteady cloud cavitation around the CAV2003 benchmark hydrofoil using passive cavitation controllers called cavitation-bubble generator (CGs) was studied. The different sizes and locations of the CGs were investigated to find a proper design of CGs to control the cloud cavitation. First, the unsteady cloud cavitation around the hydrofoil without CGs has been simulated using a Partially-averaged Navier Stokes (PANS) method to evaluate the numerical simulations based on experimental data. Second, the effect of a proper CGs as cavitation controller on the qualitative parameters such as cavitation structure and the shape of cavity was presented. The effect of passive controller on the different destructive effects of cavitation such as unsteadiness of the cloud cavitation, turbulent velocity fluctuations, high wall-pressure peaks and reduction of hydrodynamic performance has been analyzed. Our results showed that the appropriate design of the CGs on the hydrofoil surface led to the reduction in high-pressure amplitude on the wall surface of the hydrofoil. The results showed that the cyclic behavior of the cloud unsteady cavitation was suppressed partially and the hydrodynamic efficiency of the hydrofoil increased. The high wall-pressure peaks was reduced significantly. In conclusion, using this method of passive control on the surface of immersed bodies such as hydrofoil, propeller and water turbine blades in cloud cavitation regime a significant reduction in noise generation, unpleasant unsteady side force effects, high-pressure peaks on the surface, flow-induced vibration and surface erosion can be expected. The investigation of the cavitation control using passive control methods with LES simulation and experimental studies will be studied by authors in future work.

## Acknowledgement

The authors would like acknowledge Dr. Javadi from sharif university for helpful discussions. The authors would like to thank the support of computer resources by MagnitUDE supercomputers of the Center for Computational Sciences and Simulation (CCSS) at University of Duisburg-Essen.

*Corresponding Author, Ebrahim Kadivar: ebrahim.kadivar@uni-due.de

*Corresponding Author, Ebrahim Kadivar: ebrahim.kadivar@uni-due.de